\documentclass[english,aps,pre,superscriptaddress,showpacs,preprint]{revtex4-1}
\usepackage{graphicx,amsfonts,amsmath,amssymb,color}
\newcommand{\eq}[1]{$\mathrm{Eq.}$~\eqref{#1}}

\newcommand{\timeAv}[1]{\overline{#1}}
\newcommand{\ensembleAv}[1]{\langle #1 \rangle}

\newcommand{\order}[1]{\mathcal{O}(#1)}
\def\mean#1{\langle#1\rangle}
\newcommand{\Iext}{I_{\text{ext}}}

\newcommand{\Dunits}{($\mu$A/cm$^2$)$^2$\text{ms}}

\begin{document}
\title{Emergence and coherence of oscillations in star networks of stochastic excitable elements} 

\author{Justus A. Kromer}
\email{justuskr@physik.hu-berlin.de}
\affiliation{Department of Physics, Humboldt-Universit\"{a}t zu Berlin, 
Newtonstr. 15, 12489 Berlin, Germany}
\author{Lutz Schimansky-Geier}
\email{alsg@physik.hu-berlin.de}
\affiliation{Department of Physics, Humboldt-Universit\"{a}t zu Berlin, 
Newtonstr. 15, 12489 Berlin, Germany}
\affiliation{Bernstein Center for Computational Neuroscience Berlin, Germany}
\author{Alexander B. Neiman}
\email{neimana@ohio.edu}
\affiliation{Department of Physics and Astronomy and Neuroscience Program, Ohio University, Athens, Ohio 45701, USA}

\begin{abstract}
We study the emergence and coherence of stochastic oscillations in star networks of excitable elements in which peripheral nodes receive independent random inputs. 
A biophysical model of a distal branch of sensory neuron in which 
peripheral nodes of Ranvier are coupled to a central node by myelinated cable segments is used along with a generic model of networked stochastic active rotators. We show that coherent oscillations can emerge due to stochastic synchronization of peripheral nodes and that the degree of coherence can be maximized by tuning the coupling strength and the size of the network. Analytical results are obtained for the strong coupling regime of the active rotator network. In particular, we show that in the strong coupling regime the network dynamics can be described by an effective single active rotator with rescaled parameters and noise. 
\end{abstract}

\maketitle

\section{Introduction}
Spontaneous oscillations are common in biological systems on diverse
levels ranging from subcellular and single cell
\cite{Kruse2005,Kholodenko2006} up to cell networks, e.g. in the central
nervous system \cite{Buz06,Sporns2011}. In many instances, these
oscillations are noisy and can be modeled by single or coupled
stochastic excitable elements \cite{lindner2004effects} forming
complex networks \cite{Newman2003,Boccaletti2006,Arenas2008}. The
coherence of noise-induced oscillations can be maximized by tuning the
noise intensity (coherence resonance) \cite{pikovsky1997coherence}, the coupling strength (array-enhanced coherence resonance) \cite{NeiSch99, ZhoKur01}, and by varying
the number of coupled excitable elements (system size coherence resonance) \cite{toral2003system,Pikovsky2003} or the network topology
\cite{Gosak2010}.

\begin{figure}[h]
\centering
\includegraphics[width=0.5\linewidth]{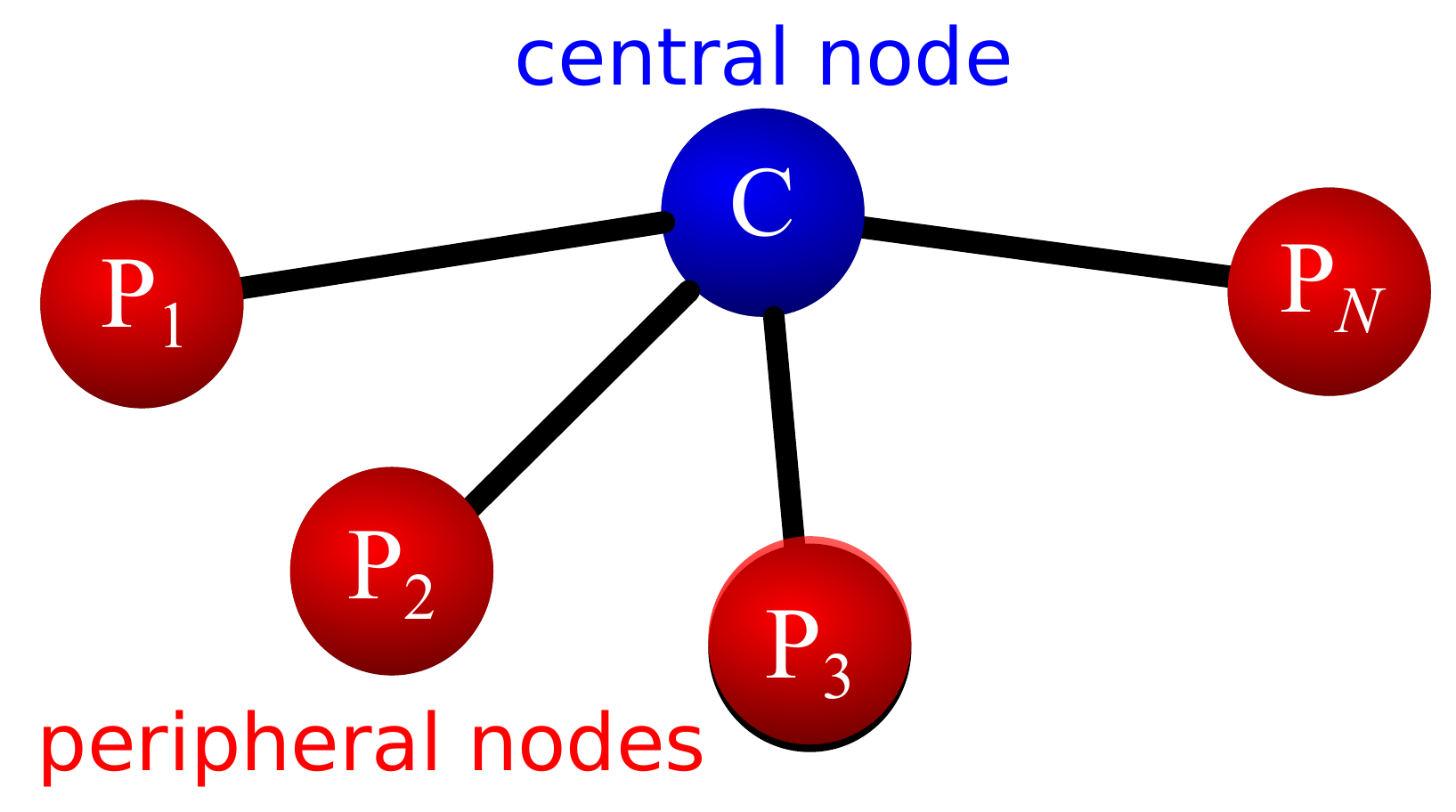}
\caption{Color online. Star network considered throughout the paper. Peripheral
  excitable elements $P_n$ are linked to the central excitable hub 
  $C$. For the model of a distal ending of the branched myelinated axon, 
  peripheral and central elements refer to the nodes of Ranvier, linked by    myelinated segments.}
\label{fig:setup}
\end{figure}

A structure, whereby few nodes, or hubs, are linked to many other
nodes, is a typical motif in topologies of scale-free random and
regular networks. A star-type topology, such as shown in
Fig.~\ref{fig:setup}, underlines a common situation when peripheral nodes
P$_n$, P$_m$, are not coupled directly, but through a central hub,
C. Several studies addressed peculiar dynamics of star networks of
{\it deterministic} oscillators
\cite{kitajima2009bifurcation,kitajima2012cluster,
  burylko2012bifurcations,kazanovich2013competition}, including the
phenomenon of remote synchronization \cite{Frasca2012,Bergner2012}.

This paper is focused on star networks of stochastic excitable
elements.  One particular example, which motivated this study, comes
from the morphology and dynamics of primary sensory neurons, which possess
branched myelinated structures in their peripheral terminals
\cite{iggo1969structure,pruszynski2014edge,lee2014sensory,bewick2015}.
Axons of such a neuron can be viewed as a tree-like cable those myelinated
segments link active excitable elements, known as nodes of Ranvier
\cite{quick1980}. Myelin terminates at the peripheral nodes, which receive sensory inputs via unmyelinated thin processes.  In such a structure, action potentials
or spikes can be initiated in several zones due to stochastic and
independent inputs to peripheral nodes, resulting in highly non-linear
interactions \cite{eagles1974afferent}. Notable examples are the
muscle spindle sensory neurons, which possess multiple spike initiation
zones \cite{quick1980} and show noisy periodic discharges \cite{banks1997pacemaker}. Drawing in Fig.~\ref{fig:setup} then
represents a distal branch of a myelinated axon where peripheral nodes P$_k$ 
receive independent stochastic inputs and are linked by myelinated
segments to the central node C, connected to the rest of the
axon. We study the emergence of spontaneous oscillations and determinants of oscillation coherence in such a structure. Two
types of models for active nodes are used for the excitable elements of the
star network. A Hodgkin-Huxley model, which we study by the use of numerical simulations, and 
an active rotator for which we perform analytical
calculations.  

The paper is organized as follows. Models and numerical
methods, as well as a brief summary of the related single node dynamics, are presented in Sec. \ref{sec:ModelsandMethods}.  Stochastic dynamics of the star network of excitable nodes of Ranvier, modeled by a Hodgkin-Huxley  type
system, is studied numerically in Sec. \ref{sec:StarNetworkHHModel}. Then, the star network of stochastic active rotators is analyzed in Sec. \ref{sec:StarNetworkARModel}.

\section{Models and methods}
\label{sec:ModelsandMethods}
In the present paper, we are interested in the stochastic dynamics of
excitable elements linked in a star motif such as the one shown in
Fig.~\ref{fig:setup}. Peripheral nodes (P$_k$) receive independent
stochastic inputs and are coupled to the central hub (C).  Random
inputs to peripheral nodes may elicit a large event or spike, which in
turn may fire up the central node. In the following, we denote the time of the $j$th spike of the central node by $t^{\text{C}}_j$  and that of the $j$th spike of 
the $n$th peripheral node by $t^{\text{P$_n$}}_j$. The sequences of spike times form corresponding spike trains. Our primary interest is the statistics
of a spike train generated at the central node. 

\subsection{Spike train statistics}

The coherence of spiking is characterized using the statistics of interspike
intervals (ISIs) $\Delta t_j^k:=t_{j+1}^k-t_j^k$, where the subscripts $k=\text{C},\text{P$_n$}$ indicate the node, e.g. central or
peripheral, and $j=1,...,J$, with $J$ being the number of spikes
generated by the $k$-th node. The mean firing rate $r^k$ of a node is
defined as reciprocal of the mean ISI and the coefficient of
variation (CV) $C_{\text{V}}^k$ is defined as the ratio of ISI standard deviation and the mean ISI, i.e.
\begin{equation}
\label{eq:rateCV}
r^k=\ensembleAv{\Delta t_j^k}^{-1}, \quad 
C_{\text{V}}^k=r^k\, \sqrt{\langle (\Delta t_j^k - \langle \Delta t_j^k  \rangle)^2 \rangle}.
\end{equation}
Averages were taken over the sequence of ISIs obtained from the spike trains of the central and the individual peripheral nodes, respectively.

\subsection{Hodgkin-Huxley type model}

We use a discrete model \cite{keener1998mathematical,ermentrout2010foundations} for an axon branch consisting of a central node of Ranvier connected to $N$
peripheral nodes by myelinated segments.  The total
current of the $k$-th peripheral node is given by
$$
\pi \mu d \left(C_{\text{n}} \dot{V_k} +I_\text{ion}\right) = i_\text{ext} + \frac{V_0-V_k}{R},
$$
where $\mu$ and $d$ are the length and diameter of the node,
respectively; $C_{\text{n}}$ is the node's capacitance per unit area;
$I_\text{ion}$ is the ionic current density; $i_\text{ext}$ the 
external current, i.e. sensory input to a peripheral node, and $R$ is
the resistance of the myelinated segment.  Similarly, for the central
node,
$$
\pi \mu d \left(C_{\text{n}} \dot{V_0} +I_\text{ion}\right) = \frac{1}{R} \sum_{k=1}^N (V_k - V_0).
$$
For a myelinated cable segment of the diameter $d$ and the length $L_m$, the resistance 
can be calculated as $R=4 L_m \rho /(\pi d^2)$, where $\rho =
200$~$\Omega\,$cm is the axoplasmic resistivity. Dividing $R$ by the
surface area of the node and taking its reciprocal provides the coupling
coefficient in units of Siemens per area: $\kappa = d/(4 \mu L_m
\rho)$.  Although we will use $\kappa$ as the control parameter, it is
useful to provide its value for a 5~$\mu$m-diameter axon with
$\mu=1$~$\mu$m long nodes of Ranvier, connected by $L_m=500$~$\mu$m
long myelinated segment: $\kappa = 125$~mS/cm$^2$.

In the following we consider a model for the node of Ranvier,
which contains only sodium and leak ionic currents. For simplicity we
consider identical nodes and so the ionic current is $I_\text{ion} =
I_\text{Na} + I_\text{L}$. For the sodium current we use the Hodgkin-Huxley (HH) type
kinetics \cite{mcintyre2002modeling}, $I_\text{Na}
= g_\text{Na} m^3 h$, where $m, h$ are activation and inactivation
variables, and $g_\text{Na}$ is the maximum value of the Na
conductance. The model's equations are:

\begin{equation}
\begin{array}{l}
C_{\text{n}}  \dot{V_0} = -g_\text{Na} m_0^3 h_0 (V_0-V_\text{Na})-g_\text{L}(V_0-V_{\text{L}}) + \kappa \, \sum_{k=1}^{N} (V_k - V_0),\\ 
C_{\text{n}}  \dot{V_k} = -g_\text{Na} m_k^3 h_k (V_k-V_\text{Na})-g_\text{L}(V_k-V_{\text{L}}) + \kappa \, (V_0 - V_k) + \Iext + \sqrt{2D} \xi_k(t),\\ 
\dot{m}_{0,k}=\alpha_m(V_{0,k}) (1-m_{0,k})- \beta_m(V_{0,k}) m_{0,k}, \\
\dot{h}_{0,k}=\alpha_h(V_{0,k}) (1-h_{0,k})- \beta_h(V_{0,k}) h_{0,k}.
\end{array}
 \label {model1}
\end{equation}
In \eq{model1} the gating kinetics of all nodes is assumed to be
identical with rate functions \cite{mcintyre2002modeling},
\begin{equation}
\begin{array}{l}
\alpha_m(V)=1.314 (V+20.4)/\{1-\exp[-(V+20.4)/10.3]\},\\
\beta_m(V)=-0.0608 (V+25.7)/\{1-\exp[(V+25.7)/9.16]\},\\
\alpha_h(V)=-0.068 (V+114)/\{1-\exp[(V+114)/11]\},\\ \beta_h(V)=2.52/\{1+\exp[-(V+31.8)/13.4]\}.
\end{array}
\label{model2}
\end{equation}
The external current applied to the peripheral nodes has a
constant component $\Iext$ and independent white noise  $\xi_k(t)$ with the intensity $D$ and
\begin{align}
\label{eq:whiteNoise1}
\langle \xi_k(t) \rangle = 0 \ \text{and} \ \langle \xi_k(t) \xi_j(t') \rangle = \delta(t-t') \delta_{kj}.
\end{align} 
Angular brackets $\langle . \rangle$ stand for the ensemble average. 
Electrical parameters of the model are
as follows $C_{\text{n}} =2~\mu$F/cm$^2$, $V_{\text{L}}=80$~mV, $V_{\text{Na}}=50$~mV,
$g_{\text{Na}}=1100$~mS/cm$^2$, $g_{\text{L}}$=20~mS/cm$^2$.  The coupling $\kappa$ 
and the constant external current $\Iext$ are the control parameters.

The coupled stochastic differential equations (\ref{model1}) were
integrated using the Euler-Maruyama method with a time step of
$10^{-4}$~ms. A spike time for a node is recorded when the voltage
crossed a threshold of $20$~mV with a positive slope. Sequences of the
spike times of the $k$-the node $\{t^k_j\}$ were recorded during
$T=1.2\times 10^6$~ms (1200s) long simulations.

\subsection{Active rotator model}

Active rotator models subject to noise are commonly used
in studies of collective dynamics in large networks of
excitable elements \cite{shinomoto1986phase, KurSch95, zaks2003noise, tessone2007theory, sonnenschein2013excitable} 
or mixtures of excitable and oscillatory elements
\cite{sonnenschein2014cooperative}.
The state of an active rotator is given by its phase $\psi$. Phase, time, and all parameters are dimensionless values. The phase dynamics can
be described as an overdamped motion in the tilted periodic potential,
\begin{equation}
  \label{eq:tilded}
  U(\psi) = -\omega \psi + V(\psi), 
\end{equation}
where $\omega$ represents the tilt, which can be considered as an
input to the rotator, and $V(\psi)$ is $2\pi$ periodic. 
There exists a critical tilt $\omega_c$ at which the extrema
of $U(\psi)$ vanish, corresponding to a saddle-node bifurcation. The
rotator is excitable for $\omega < \omega_c$, that is, an additional
input, e.g. noise, is required for the rotator to overcome the
potential barrier and generate an event. The oscillatory regime occurs
when $\omega > \omega_c$.

In the present paper, we consider two particular choices for $V(\psi)$: the
cosine potential $V_{\text{cos}}(\psi)$ and a potential $V_\text{opt}(\psi)$ introduced in Ref. \cite{LinKos01}, which provides more coherent sequences of events,
\begin{equation}
V_{\text{cos}}(\psi) =  - \cos \psi, \qquad V_\text{opt}(\psi) = \frac{\Delta}{\epsilon}\exp\left[\epsilon(1 - \cos \psi)\right].
\label{potentials.eq}
\end{equation}
The potential $V_\text{opt}$ approaches the cosine potential for $\epsilon
\to 0$, whereas large values of $\epsilon \gg 1$ yield a delta-like
potential barrier.  The parameter $\Delta$ scales the barrier height and is set to
\begin{equation}
  \Delta = \left[ \exp \left( \epsilon - \frac{1}{2} + 
    \sqrt{\epsilon^2 + \frac{1}{4}} \right)\sqrt{1-\frac{1}{\epsilon^2} \left( \frac{1}{2} - \sqrt{\epsilon^2 + \frac{1}{4}} \right)^2} \right]^{-1},
\end{equation}
which rescales the maximum value of $V_{\text{opt}}'(x)$ to one. With this normalization both $V_\mathrm{cos}$ and $V_\mathrm{opt}$ potentials yield the transition from the 
excitable to oscillatory regime at the critical tilt $\omega_c=1$.

The active rotators are coupled by a sine function of the phase differences, which is often used in networks of active 
rotators \cite{shinomoto1986phase, zaks2003noise, sonnenschein2013excitable, sonnenschein2014cooperative}. 
We distinguish between the central node with phase $\theta$ and a set
of $N$ peripheral nodes with phases $\phi_n$, $n=1,...,N$.
The equations for the central ($\theta$) and peripheral ($\phi_n$)
nodes are,
\begin{eqnarray}
&&\dot{\theta}=\omega_\theta + G(\theta)  + \kappa \sum \limits_{n=1}^{N} \sin(\phi_n- \theta)  + \sqrt{2 D_\theta} \, \xi_{\theta}(t), \nonumber \\
&&\dot{\phi_n}=\omega_{\phi_n}+ G(\phi_n) + \kappa \sin(\theta-\phi_n) + \sqrt{2 D_{\phi_n}} \, \xi_{\phi_n}(t),
\label{eq:NetworkDynamics}
\end{eqnarray}
where $G(x) = -V_{\text{cos}}'(x)$ or $G(x) = -V_{\text{opt}}'(x)$. The $\theta$-node is parametrized by 
$0 < \omega_{\theta}<1$ and the coupling strength $\kappa$. The 
$n$-th peripheral node is parametrized by the driving parameters $\omega_{\phi_n}$ and the coupling strength $\kappa$.
Furthermore, peripheral nodes are excited by independent Gaussian white noises with intensities $D_{\phi_n}$. The Gaussian noises $\xi_k(t)$ are defined as in  \eq{eq:whiteNoise1} where $j,k=\phi_n,\theta$. We also introduce
noise to the central node, but in order to model the situation in the branched myelinated axon, we will assume that its intensity is much
smaller than those of the driver noises in peripheral nodes, $D_\theta \ll
D_{\phi_n}$. Nevertheless, our analytical results are valid for arbitrary choices of the noise intensities. 

An active rotator generates an event or spike when its state variable
($\phi_n$ or $\theta$) crosses 2$\pi$ with a positive slope. We
perform a phase reset after each $2 \pi$-crossing by setting the
respective phase to zero.  Numerically, the reset is realized by
subtracting $2 \pi$ from $\phi$ whenever $\phi$ is larger than $2
\pi$.  This prevents the generation of spikes due to a crossing of $2
\pi$ with negative phase velocity. Note that this setup allows for
phases in the range $(-\infty,2 \pi)$.  

Numerical integration of
Eqs.~(\ref{eq:NetworkDynamics}) was performed using the Euler-Maruyama
method. The integration time step was set to the minimum value of  $10^{-3}/(D_{\phi_n} \kappa)$ and $5\times 10^{-3}$. Simulations were 
stopped after $10^6$ spikes of the central and peripheral nodes were detected or a maximum simulation time of $10^8$ was reached.

Statistics of spiking events generated by a single stochastic rotator,
\begin{equation}
\dot \psi = \omega - V'(\psi) + \sqrt{2D} \,\xi(t).
\label{singlerot.eq}
\end{equation}
can be calculated analytically.
In particular, an ISI corresponds to the first passage of $2 \pi$ when $\psi$ has
initially started at zero.  For the one-dimensional Brownian motion described by Eq. (\ref{singlerot.eq}) an iterative scheme has been developed in
Refs. \cite{PonAnd33,siegert1951first} and, in case of a tilted
periodic potential, simplified formulas for the mean first passage time
(corresponding to the mean ISI) and the variance of the first passage
time distribution (corresponding to the variance of ISIs) have been
developed in Refs.~\cite{LinKos01, reimann2002diffusion}.  Applying
these results to \eq{singlerot.eq} with a spike generated at
$\psi=2 \pi$ and an instantaneous phase reset to $\psi=0$, the mean ISI is
given by
\begin{equation}
\label{eq:SelfConstmeanISI}
\mean{\Delta t} = 
\frac{
\int \limits_{0}^{2 \pi}dx \int \limits_{x-2 \pi}^{x}dy \, \frac{\Phi(x)}{\Phi(y)}}
{D \left( 1- e^{-\frac{2\pi \omega }{D}}\right)},
\end{equation}  
and the variance of ISIs by
\begin{equation}
\label{eq:VarianceStart}
\text{var}(\Delta t)= 
\frac{2 \int \limits_0^{2 \pi}dx \  \left( \int \limits_{x-2\pi}^{x} dy \ \Phi(y)^{-1} \right)^2 \int \limits_x^{x+ 2 \pi}dz \ \Phi(z)\Phi(x) }{D^2\left(1-e^{-\frac{2 \pi \omega}{D}}\right)^3},\\
\end{equation}
where $\Phi(x)=\exp[U(x)/D]$.
\begin{figure}[h!]
  \centerline{\includegraphics[width=0.75\textwidth]{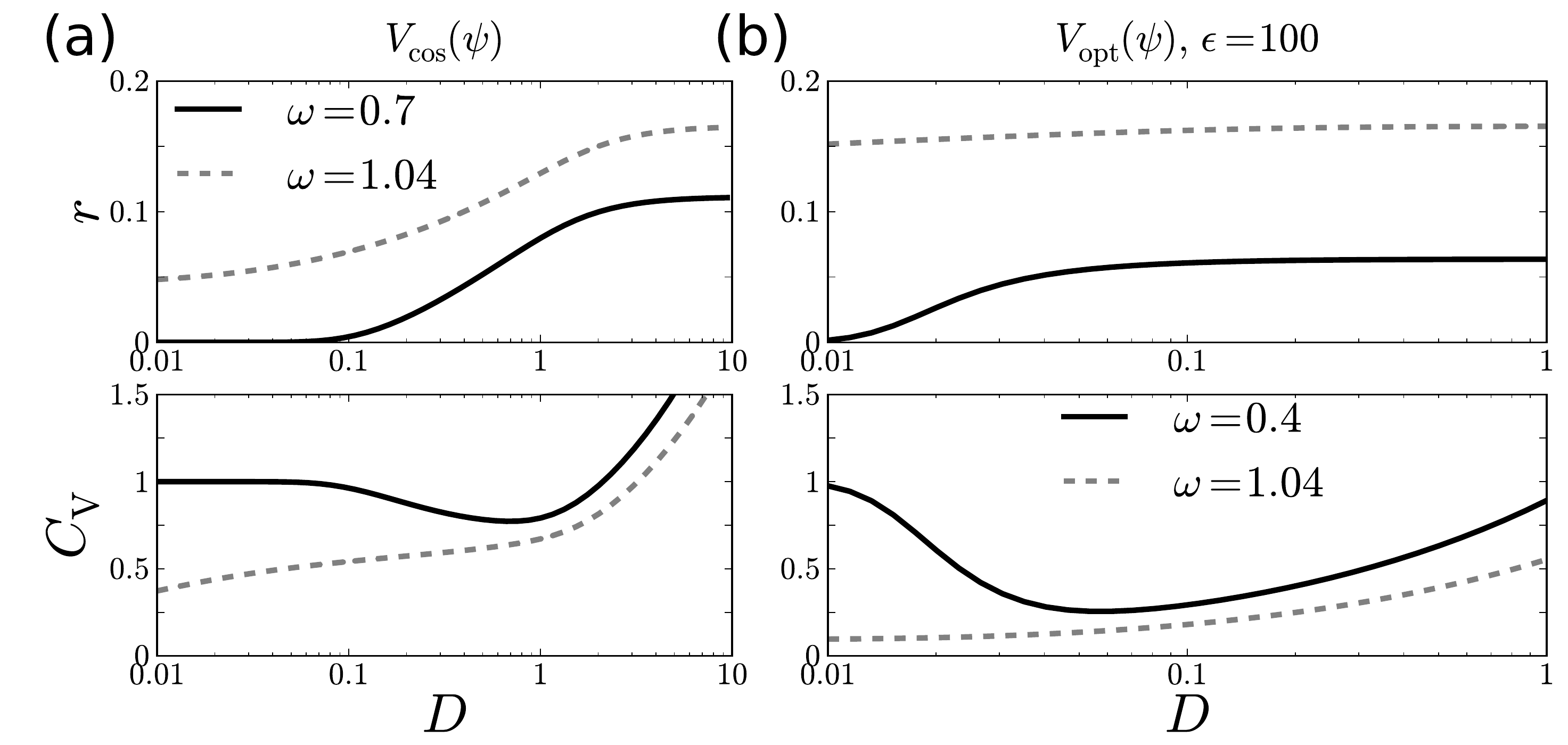}}
  \caption{Dependence of the mean firing rate $r$ (top) and $C_V$ (bottom) on noise
    intensity $D$ for different values of $\omega$, of a single active rotator, Eq. (\ref{singlerot.eq}), 
    for the two potentials,  Eq. (\ref{potentials.eq}), (a) and (b) with
   indicated value of $\epsilon$, respectively. 
    }
  \label{fig:single_rotator}
\end{figure}
We use these formulas to calculate the firing rate $r = 1/\mean{\Delta t}$ and the CV $C_\mathrm{V}= \sqrt{\text{var}(\Delta t)}/\mean{\Delta t}$, which is in agreement with Eq.~(\ref{eq:rateCV}).  
Figure \ref{fig:single_rotator} shows the mean firing rate and the CV for $\omega < \omega_c$, i.e. in the excitable regime, 
and for one $\omega > \omega_c$, i.e. in the oscillatory regime, respectively for 
both potentials used in this study as functions of the noise intensity. 
In the excitable regime, the 
CV passes through a minimum for both potentials, demonstrating the
phenomenon of coherence resonance \cite{pikovsky1997coherence}. As
expected, noise-induced oscillations are more coherent for the 
potential $V_\text{opt}$. In the oscillatory regime, noise merely degrades the oscillation coherence, as CV increases monotonically with $D$.

\section{Results}

\subsection{Noise-induced spiking in a star network with Hodgkin-Huxley nodes}
\label{sec:StarNetworkHHModel}
We start with the deterministic dynamics ($D=0$) of a single uncoupled
peripheral node, i.e. $N=1$, $\kappa=0$ in Eqs.~(\ref{model1}).  For $\Iext
< 29.06$~$\mu$A/cm$^2$, the node is at the stable resting equilibrium. This equilibrium undergoes a subcritical Andronov-Hopf
bifurcation at $\Iext \approx 29.06$~$\mu$A/cm$^2$. Oscillations on the large-amplitude
limit cycle cause a periodic sequence of full-size action
potentials for $\Iext > 29.06$~$\mu$A/cm$^2$.
\begin{figure}[h!]
  \centerline{\includegraphics[width=0.75\textwidth]{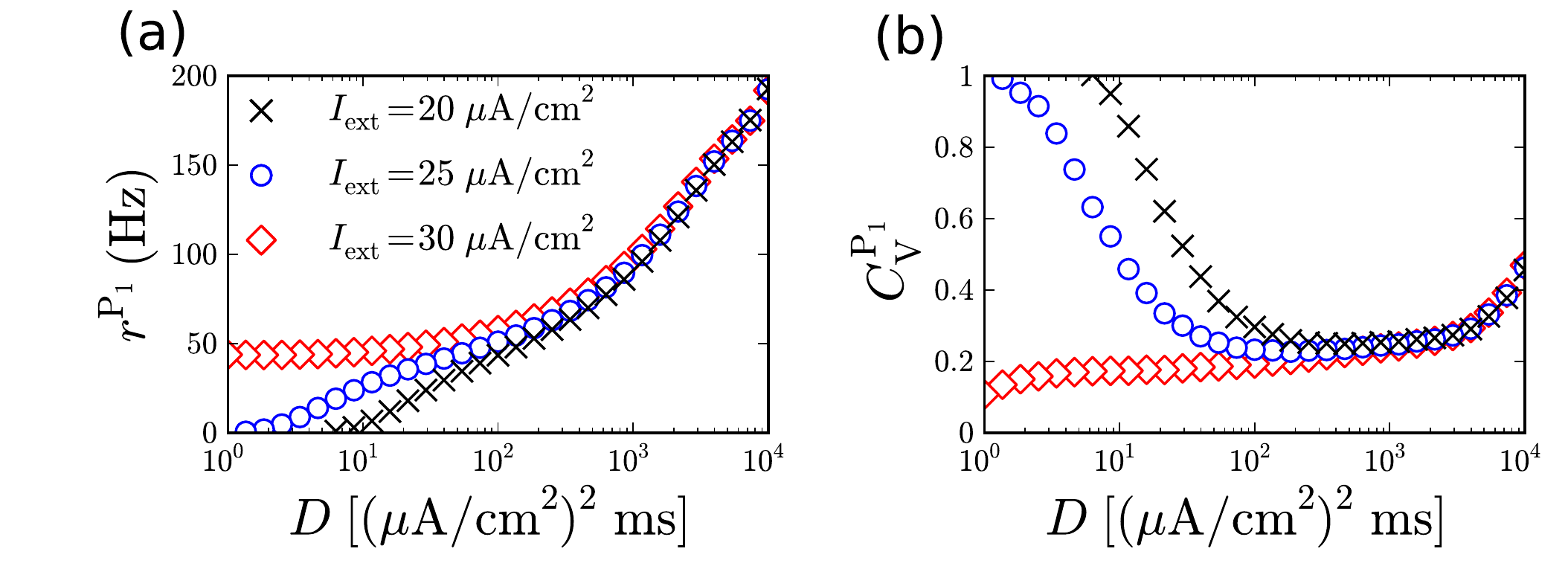}}
  \caption{Color online. Dependence of the mean firing rate $r^{\text{P}_1}$ and $C_ {\text{V}}^{\text{P}_1}$ on  noise intensity $D$ of a single uncoupled peripheral HH-type node for
    the indicated values of $\Iext$.  
}
  \label{stoch-1node.fig}
\end{figure}

With the addition of stochastic input the single node exhibits the
phenomenon of coherence resonance \cite{pikovsky1997coherence} in the
excitable regime, $\Iext < 29.06$~$\mu$A/cm$^2$, demonstrated in
Fig.~\ref{stoch-1node.fig}. For $\Iext < 29.06$~$\mu$A/cm$^2$ and weak
noise, the node  generates action potentials extremely rarely, resulting in a low firing rate and a large CV (black and blue symbols in
Fig.~\ref{stoch-1node.fig}a,b). With the increase of the noise intensity, the firing rate increases and the ISI sequence becomes progressively more regular, as
indicated by the decrease of the CV. Strong noise eventually results in an
increase of the CV.  In contrast, in the regime of periodic spiking,
$\Iext > 29.06$ $\mu$A/cm$^2$, the CV increases monotonically with $D$,
(red squares in Fig.~\ref{stoch-1node.fig}b), as noise merely
worsens the coherence of periodic spiking.

\begin{figure}[h!]
  \centerline{\includegraphics[width=0.9\textwidth]{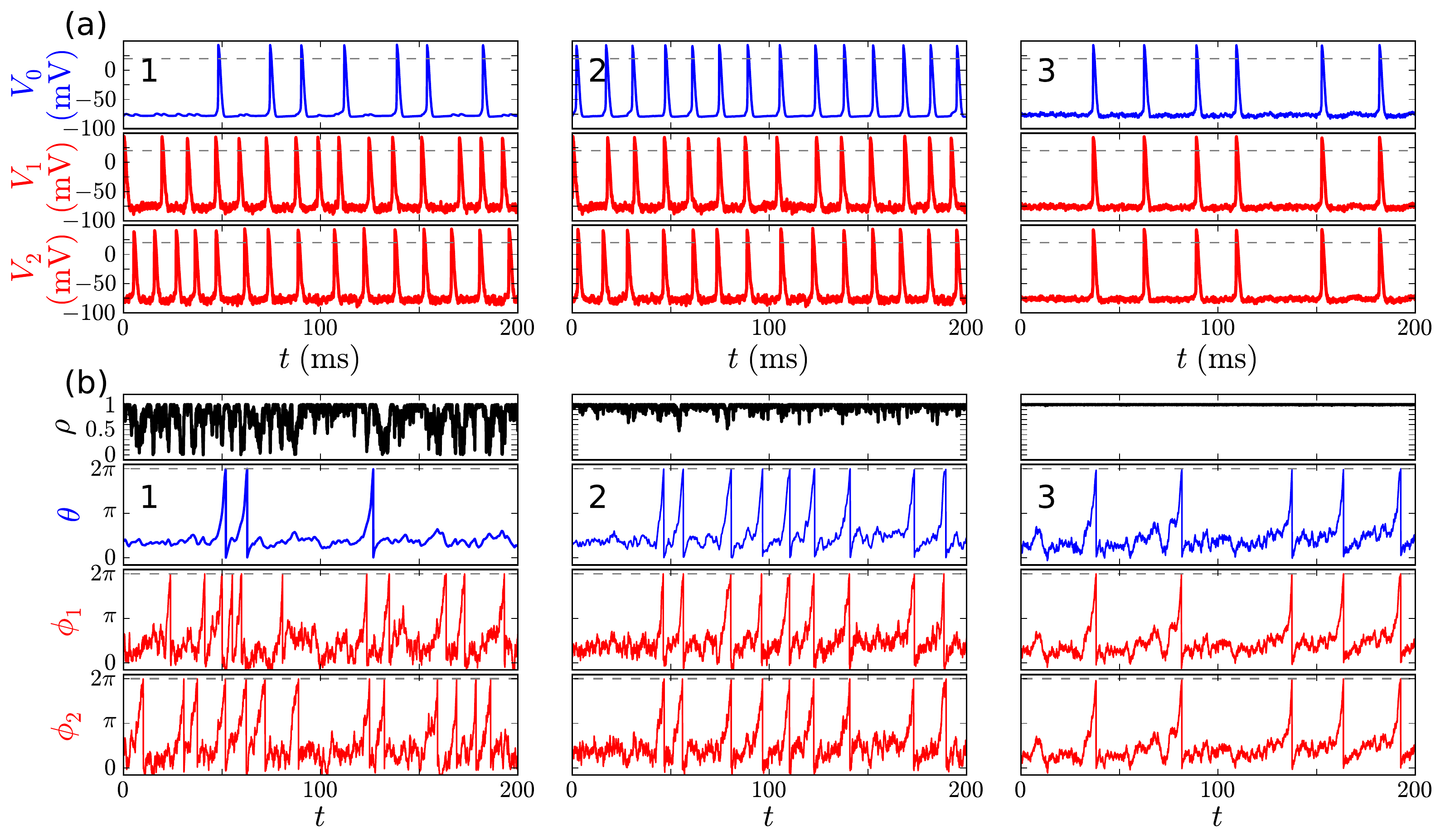}}
  \caption{Color online. State traces of  a star network of HH (a) and active rotator (b) stochastic excitable elements for different values of the coupling strengths, $\kappa$. The central node is plotted blue and peripherals red, respectively. Spiking thresholds are indicate by the dashed horizontal lines.
(a): Voltage traces of a star network with HH nodes, Eqs. (\ref{model1}) with $N=2$ peripheral nodes. The parameters are:
$\kappa=0.3$ (1), $\kappa=0.625$ (2), $\kappa=100$~mS/cm$^2$ (3), respectively;  $\Iext = 20$ $\mu$A/cm$^2$, $D=500$~\Dunits.
(b): Traces  of the phases  in a similar network of active rotators, \eq{eq:NetworkDynamics}, with $G(x)=-V_{\text{cos}}'(x)$.  
The trace of the Kuramoto order parameter, \eq{eq:KuramotoOrder}, is illustrated by the black line. The parameters are:   
$\kappa=0.328$ (1), $\kappa=2.147$ (2), $\kappa = 57.646$ (3), respectively;
$D_{\theta}=0$, $D_{\phi_n}=D_{\phi_m}=0.4$, $\omega_{\theta}=\omega_{\phi_n}=\omega_{\phi_m}=0.9$.
The average degrees of synchronization $\overline{\rho}$, Eq. (\ref{eq:timeAvRhoDefinition}),  are $0.78$ (1), $0.95$ (2), and $1.0$ (3), respectively.}
  \label{stoch-traces.fig}
\end{figure}

\begin{figure}[h!]
  \centerline{\includegraphics[width=0.75\textwidth]{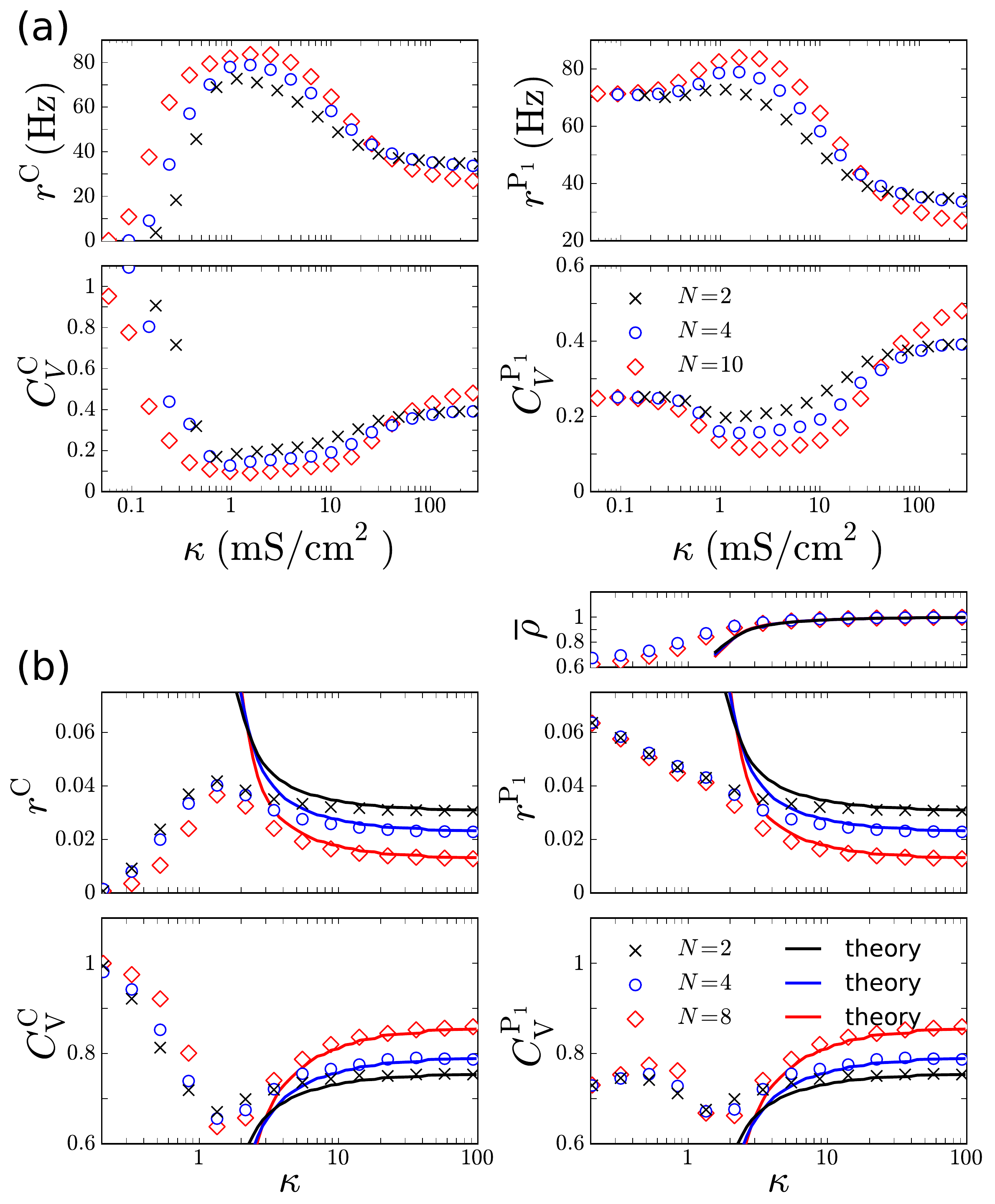}}
  \caption{Color online. Effect of coupling on the firing statistics of star networks with
HH (a) and active rotator (b) nodes. Firing rates $r$ and CVs $C_\mathrm{V}$ versus the coupling strength $\kappa$ for the indicated numbers of peripheral nodes $N$. Left columns correspond to the central node and right columns to the first peripheral node, respectively.
(a): Star network of HH-type nodes. The parameters are:
$\Iext = 20$~$\mu$A/cm$^2$ and $D=500$~\Dunits.
(b): Star network of active rotators, Eq. (\ref{eq:NetworkDynamics}), with cosine potential, $G(x)=-V_{\text{cos}}'(x)$. The upper right panel shows the time-averaged order parameter $\timeAv{\rho}$. Symbols correspond to numerical simulation, solid lines show analytical approximation. The parameters are:
 $\omega_{\phi_n}=\omega_{\phi_m}=\omega_{\theta}=0.9$, $D_{\theta}=0$, and $D_{\phi_n}=D_{\phi_m}=0.4$.}
  \label{stoch-vs-cpl.fig}
\end{figure}

When $N$ excitable nodes are coupled in the star network, stochastic
dynamics of the central node depends crucially on the coupling
strength.  For extremely weak coupling, peripheral nodes fire asynchronously. This results in sparse and irregular firing of the
central node, see Fig.~\ref{stoch-traces.fig}a1. Weak coupling
synchronizes the network leading to coherent spiking, see 
Fig.~\ref{stoch-traces.fig}a2. Finally, strong coupling makes the
system stiff, so that inputs to peripheral nodes may not be
enough to sustain coherent periodic firing. Consequently, nodes fire in
synchrony, but less coherently, as Fig.~\ref{stoch-traces.fig}a3 indicates.
This dynamics is summarized in Fig.~\ref{stoch-vs-cpl.fig}a. The main
feature is the existence of an optimal value of the coupling strength, 
which maximizes the firing rate of peripheral and the central nodes
and minimizes the variability of their firing.

\begin{figure}[h!]
  \centerline{\includegraphics[width=0.9\textwidth]{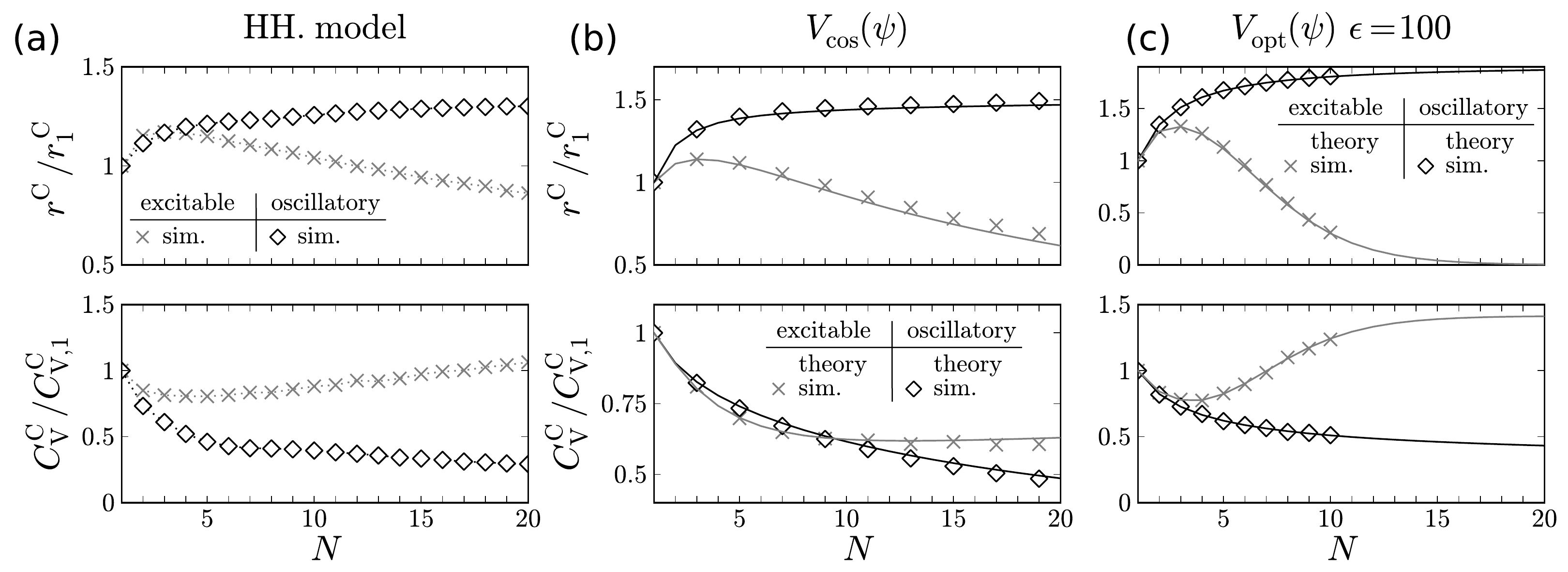}}
  \caption{System size effect on the firing statistics of star networks of HH-type (a) and active rotator nodes (b,c) in the strong coupling regime. 
Upper panels show the firing rate of the central node and lower panels show the CV versus the number of peripheral nodes.
Two regimes of peripheral nodes, excitable and oscillatory, are shown.
Firing rates and CVs are normalized by their values at $N=1$, $r^{\text{C}}_1$ and $C_{\text{V},1}^{\text{C}}$, respectively, in order to show the relative change of these measures.  
  (a): Star network of HH-type nodes. Symbols show results of numerical simulations. The parameters are:   
$I_{\text{ext}}=1$~$\mu$A/cm$^2$ (excitable regime) and $I_{\text{ext}}=60$~$\mu$A/cm$^2$ (oscillatory regime),  $D=500$~\Dunits.
(b): Star network of active rotators with cosine potential, $V_\mathrm{cos}$.
The parameters are: $\omega_{\theta}=0.3$, $D_{\theta}=0$, $\kappa=100$; $\omega_{\phi_n}=\omega_{\phi_m}=0.7$, $D_{\phi_n}=D_{\phi_m}=10$ (excitable regime) and  $\omega_{\phi_n}=\omega_{\phi_m}=1.5$, $D_{\phi_n}=D_{\phi_m}=5$ (oscillatory regime).
(c): Star network of active rotators with potential $V_\mathrm{opt}$.
The parameters are: $\omega_{\theta}=0$, $D_{\theta}=0$, $\kappa=1000$, $D_{\phi_n}=D_{\phi_m}=0.3$; $\omega_{\phi_n}=\omega_{\phi_m}=0.1$ (excitable regime) and  $\omega_{\phi_n}=\omega_{\phi_m}=1.1$ (oscillatory regime). On panels b,c symbols correspond to numerical simulations and lines show theory.
}
  \label{stoch-vs-N.fig}
\end{figure}

Another feature is concerned with the scaling with the
number of peripheral nodes $N$ for the case of  strong coupling, which is illustrated in Fig.~\ref{stoch-vs-N.fig}a.  The strong coupling regime is of particular interest as it represents the situation 
for primary endings of muscle spindles sensory neurons. For example, the length of myelinated segments in peripheral branches of cat muscle spindles ranges over 58 -- 192 $\mu$m \cite{banks1982form}, 
which for a cable diameter of 1 -- 6 $\mu$m corresponds to the coupling strength $\kappa=$ 65 -- 1300~mS/cm$^2$, causes strong coupling between central and peripheral nodes. 

In a small star network, $N<3$, the variability of the spiking 
of the central node is suppressed by the increase
of the number peripheral nodes as indicated by smaller values of $C_{\text{V}}$ in
Fig.~\ref{stoch-vs-N.fig}a (bottom). On
the contrary, the operation mode of the peripheral nodes is critical for the scaling for larger $N$ and strong coupling. 
If peripheral nodes are excitable, the dependencies of $r^{\text{C}}$ and $C_{\text{V}}^{\text{P}}$ on $N$ are non-monotonic. In that case,
we find an optimal network size for which the firing rate attains a maximum and the spiking is most coherent. Thus, in the
biologically-plausible strong coupling regime, the network
demonstrates system size resonance \cite{toral2003system,Pikovsky2003}.
In contrast, if peripherals operate in the oscillatory regime, the firing rate increases, while the variability decreases monotonically. 

\subsection{Star network of active rotators}
\label{sec:StarNetworkARModel}

In this section we show that in the strong coupling limit the star network of stochastic active rotators can be replaced with a single effective rotator with an effective drive and noise.
However, first we illustrate that the dynamics of the star network of active rotators is qualitatively akin to that of networked HH nodes.
Figure~\ref{stoch-traces.fig}b shows the effect of regularization of the nodes' firing for an optimal coupling strength. This can be also seen in Fig.~\ref{stoch-vs-cpl.fig}b, which shows the firing rate and CV versus $\kappa$ for the central node and peripherals. 
Similar to the star network of HH nodes, the firing rate passes through a maximum and the CV through a minimum when peripherals are in excitable regime. Furthermore, in case of strong coupling, the network shows the effect of system size resonance for both potentials used, as shown in Fig.~\ref{stoch-vs-N.fig}b,c.

Next, we study the spiking in the strong coupling regime, corresponding to a model of a branched myelinated axon.  Iin that case, as illustrated in Fig.~\ref{stoch-traces.fig}, the membrane potentials in the HH network as well as the phases in the active rotator network, are synchronized.
In order to quantify the degree of synchronization of the peripheral nodes, we use the time-dependent Kuramoto order parameter:
\begin{eqnarray}
\label{eq:KuramotoOrder}
 \rho(t) \ e^{I \Psi(t)}:=\frac{1}{N} \sum \limits_{i=1}^{N} e^{I \phi_i}.
\end{eqnarray}
Its absolute value $\rho(t)$, yields the degree of
synchronization of all peripheral nodes, i.e. $\rho(t)=1$ corresponds to perfect synchronization of the peripherals, while $\rho(t)=0$ indicates the completely
asynchronous regime. $\Psi(t)$ is the peripherals' mean field phase.
Using the definition of the order parameter, Eq.~(\ref{eq:KuramotoOrder}), and the $\theta$-dynamics, Eq.~(\ref{eq:NetworkDynamics}), we obtain for the central node,
\begin{align}
  \label{eq:ThetaDynamicsMeanField}
  \dot{\theta}={\omega_{\theta}}+G(\theta) + N \rho
  {\kappa} \sin(\Psi- \theta) + \sqrt{2 D_{\theta}}
  \xi_{\theta}(t).
\end{align}
Thus, the central node is only coupled to the phase of the mean field  of the peripherals $\Psi$ with the rescaled coupling strength $N \rho(t)
\kappa$. Consequently, the central node is most affected if the peripheral nodes are synchronized.

The magnitude and the phase of the order parameter can be obtained by following an 
approach of Ref.~\cite{tessone2007theory}.
Taking the time derivative of the order parameter, Eq. (\ref{eq:KuramotoOrder}),
\begin{align}
 \frac{d}{dt}\left( \rho e^{I \Psi} \right)=\dot{\rho} e^{I \Psi} + I \dot{\Psi} \rho e^{I \Psi} = \frac{I}{N} \sum \limits_{i=1}^{N} \dot{\phi_i} e^{I \phi_i}\notag,
\end{align}
and dividing both sides by $e^{I \Psi}$, we obtain for the imaginary part 
\begin{align}
\label{eq:EquationForPsiStep1}
 \rho  \dot{\Psi} = \frac{1}{N} \sum \limits_{i=1}^{N} \dot{\phi_i} \cos \left(\phi_i - \Psi \right).
\end{align}
Here, the phase difference $\delta_i^{\Psi}:= \phi_i-\Psi$ between the
$i$-th peripheral node and the mean field appears, which we split into two parts.
First, the phase difference between the mean field and the phase of the central node $\delta^{\theta,\Psi}:=\theta - \Psi$ and, second, the  
differences between the phase of the central node and those of the $N$ peripheral nodes $\delta_i^{\theta}:=\theta-\phi_i$. It holds that 
$\delta_i^{\Psi}=-\delta_i^{\theta}+ \delta^{\theta,\Psi}$. 
For strong coupling,  $N \rho \kappa \gg \text{max}(1,\omega_{\theta},D_{\theta})$ and $\kappa \gg \text{max}(1,\omega_{\phi_n},D_{\phi_n})$,
both phase differences, $\delta^{\theta,\Psi}$ and $\delta_i^{\theta}$ become small parameters and so the cosine in Eq.~(\ref{eq:EquationForPsiStep1}) can be linearized, yielding
\begin{align}
\label{eq:EquationForPsiStep2}
 \rho  \dot{\Psi} = \frac{1}{N} \sum \limits_{i=1}^{N} \dot{\phi_i} + \order{(\delta_i^{\Psi})^2}.
\end{align}
Using Eq.~(\ref{eq:NetworkDynamics}), and the
definition of the Kuramoto order parameter, Eq. (\ref{eq:KuramotoOrder}), 
we obtain in the first order  in $\delta_i^{\Psi}$  
\begin{align}
  \label{eq:PsiDynamicsStrongCoupling}
  \rho \dot{\Psi} = \langle \omega_{\phi}\rangle_N + \langle G(\phi_i)\rangle_N + \kappa \rho \sin(\theta - \Psi) + \sqrt{\frac{2 \langle D_{\phi}\rangle_N}{N}} \xi_{\Psi}(t),
\end{align}
where $\langle . \rangle_N$ denotes averages over all peripheral nodes, i.e.  $\langle D_{\phi} \rangle_N:=\frac{1}{N} \sum_{n=1}^N D_{\phi_n}$ and $\langle \omega_{\phi} \rangle_N:=\frac{1}{N} \sum_{n=1}^N \omega_{\phi_n}$.
In Eq.~(\ref{eq:PsiDynamicsStrongCoupling}), $\xi_{\Psi}(t)$ is a white Gaussian noise, which has an intensity that is inversely 
proportional to the network size and
proportional to the average noise intensity of the peripheral nodes \cite{tessone2007theory}.

If $G(x)=-V_{\text{cos}}'(x)$, the average of $G(x)$ can be simplified using the definition of the Kuramoto order parameter, \eq{eq:KuramotoOrder},
yielding similar results as in \cite{tessone2007theory},  
\begin{align}
  \label{eq:PsiDynamicsStrongCouplingV}
  \rho \dot{\Psi} = \langle \omega_{\phi}\rangle_N - \rho \sin(\Psi) + \kappa \rho \sin(\theta - \Psi) + \sqrt{\frac{2 \langle D_{\phi}\rangle_N}{N}} \xi_{\Psi}(t).
\end{align}
Small deviations from the perfectly-synchronized state can be accounted for by representing $\rho$ by its long-time average,
\begin{eqnarray}
\label{eq:timeAvRhoDefinition}
\timeAv{\rho}:=\frac{1}{T} \int \limits_{0}^T \ dt' \ \rho(t),
\end{eqnarray}  
where  $T \gg 1/r^{i}$, $r^{i}$ are the mean firing rates of peripheral and central nodes, with $i = \text{C}, \text{P}_1, \text{P}_2,..,\text{P}_N$. 
This results in the strong coupling approximation for the mean field phase,
\begin{align}
  \label{eq:PsiDynamicsStrongCouplingVTimeAverage}
   \dot{\Psi} = \frac{\langle \omega_{\phi}\rangle_N}{\timeAv{\rho}} - \sin(\Psi) + \kappa \sin(\theta - \Psi) + \sqrt{\frac{2 \langle D_{\phi}\rangle_N}{\timeAv{\rho}^2 N}} \xi_{\Psi}(t).
\end{align}
For an arbitrary potential, an equation of similar form can be obtained by considering only the zeroth order contributions of $\delta_i^{\Psi}$ in \eq{eq:PsiDynamicsStrongCoupling}, i.e. $\phi_i=\Psi$ and
$\rho = 1$. This yields
\begin{align}
  \label{eq:PsiDynamicsStrongCouplingGeneralTimeAverage}
   \dot{\Psi} = \langle \omega_{\phi}\rangle_N + G(\Psi) + \kappa \sin(\theta - \Psi) + \sqrt{\frac{2 \langle D_{\phi}\rangle_N}{N}} \xi_{\Psi}(t), \ \ \ \rho =1.
\end{align}
In either case, the dynamics of the strongly-coupled star network
is described by two coupled rotators in which the central node, \eq{eq:ThetaDynamicsMeanField}, is coupled to a second rotator, which represents the dynamics of the mean field phase of the peripheral nodes. In the case of cosine potential  this two-rotator representation reads,
\begin{eqnarray}
  \label{eq:TwoOscillatorRepresentation}
  \dot{\theta}&=&\omega_{\theta}-\sin(\theta) + N \timeAv{\rho} \kappa \sin(\Psi- \theta)  + \sqrt{2 D_{\theta}} \xi_{\theta}(t),\\
  \dot{\Psi} &=& \frac{\langle \omega_{\phi}\rangle_N}{\timeAv{\rho}} - \sin(\Psi) + \kappa \sin(\theta - \Psi) + \sqrt{\frac{2 \langle D_{\phi}\rangle_N }{\timeAv{\rho}^2 N}} \xi_{\Psi}(t).\nonumber
\end{eqnarray}

In order to obtain analytical results for the ISI statistics of the central
node, we consider the dynamics of the phase difference
$\delta^{\theta,\Psi}(t) = \theta(t)-\Psi(t)$, which is small in the strong
coupling regime, $\delta^{\theta,\Psi}(t) \ll 1$. Hence, linearizing the coupling
functions in Eq.~(\ref{eq:TwoOscillatorRepresentation}) yields,
\begin{eqnarray}
\label{eq:ThetaFirstOrder}
  \dot{\theta} &=& \omega_{\theta} - \sin ( \theta )- N \timeAv{\rho} \kappa \ \delta^{\theta,\Psi} + \sqrt{2 D_{\theta}} \xi_{\theta}(t),\\
  \dot{\Psi} &=& \frac{\langle \omega_{\phi}\rangle_N}{\timeAv{\rho}} - \sin(\Psi) + \kappa \ \delta^{\theta,\Psi} + \sqrt{\frac{2 \langle D_{\phi}\rangle_N}{\timeAv{\rho}^2 N}} \xi_{\Psi}(t).\nonumber
\end{eqnarray}
Furthermore, as we show in Appendix \ref{ap:calculationDelta}, the phase difference $\delta^{\theta,\Psi}(t)$ can be approximated by 
\begin{align}
\label{eq:OU}
\delta^{\theta,\Psi}(t) \approx 
\frac{\omega_1}{\kappa_1}+\frac{\sqrt{2D_1}}{\kappa_1}\,\xi_1(t),
\end{align}
where $\omega_1=\omega_{\theta} - \langle \omega_{\phi}\rangle_N / \timeAv{\rho}$,
$\kappa_1=\kappa \left(N \timeAv{\rho} + 1 \right)$, and $\xi_1(t)$ is a Gaussian noise, given by
\begin{align}
\label{eq:Xi1Noise}
\sqrt{2 D_1} \xi_1(t)=\sqrt{2 D_{\theta}} \xi_{\theta}(t)-\sqrt{\frac{2 \langle D_{\phi}\rangle_N}{\timeAv{\rho}^2 N}} \xi_{\Psi}(t),
\end{align}
with $D_1=D_{\theta}+\langle D_{\phi}\rangle_N/(N \timeAv{\rho}^2)$. 
Putting the approximation for $\delta^{\theta,\Psi}(t)$ into \eq{eq:ThetaFirstOrder} with (\ref{eq:Xi1Noise}), we obtain an effective single-rotator
approximation for the star network with rescaled driving parameter and noise intensity,
\begin{align}
\label{eq:ThetaStrongCoupling}
\dot{\theta} = \omega_{\text{mod}}(\timeAv{\rho},N) - \sin ( \theta )+ \sqrt{2
  D_{\text{mod}}(\timeAv{\rho},N)}\, \xi(t).
\end{align}
The first term on the r.h.s. stands for the effective drive
\begin{align}
\label{eq:modulateddrive}
\omega_{\text{mod}}(\timeAv{\rho},N)= \frac{\langle \omega_{\phi}\rangle_N}{\timeAv{\rho}}+\frac{1}{1 + N \timeAv{\rho}} \left(\omega_{\theta}-\frac{\langle \omega_{\phi}\rangle_N}{\timeAv{\rho}}\right).
\end{align}
The effective noise intensity $D_{\text{mod}}(\timeAv{\rho},N)$ accounts for the impact of white Gaussian noises of all nodes in the network and is given by
\begin{align}  
\label{eq:modulatednoise}
D_{\text{mod}}(\timeAv{\rho},N) = \frac{D_{\theta} + N \langle D_{\phi}\rangle_N}{\left( 1 + N \timeAv{\rho}\right)^2}.
\end{align}
Note that the dependence on the coupling strength $\kappa$ is captured by 
the average degree of synchronization $\timeAv{\rho}$. The latter can be calculated numerically using the self-consistent approach presented in 
Appendix \ref{ap:calculationRho}.
In the general case of arbitrary 
functions $G(x)$, one obtains a similar approximation in the limit of perfectly-synchronized peripherals, $\timeAv{\rho} = 1$, which reads
\begin{eqnarray}
\label{eq:ArbitraryG}
\dot{\theta} = \omega_{\text{mod}}(1,N) + G(\theta)+ \sqrt{2 D_{\text{mod}}(1,N)} \xi(t).
\end{eqnarray}
Note that this result does not account for the influence of the coupling strength and solely holds in the limit $\kappa \rightarrow \infty$.
We remark that the effective dynamics of the mean field phase of the peripheral nodes $\Psi$, is described by an identical equation.  

\begin{figure}[h!]
\centerline{\includegraphics[width=0.75\textwidth]{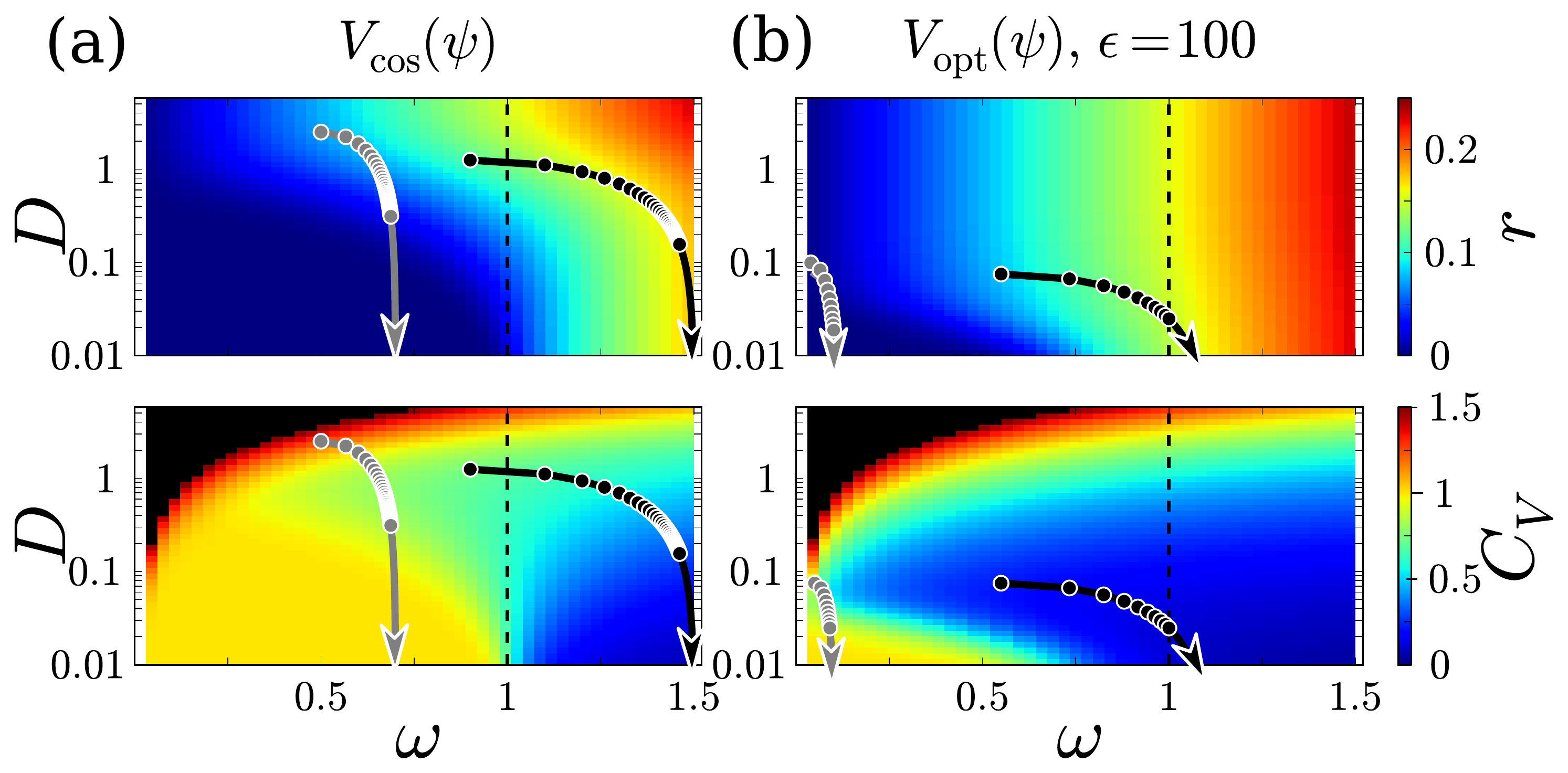}}
\caption{Color online. 
Dependence of the mean firing rate $r$ and $C_V$ on noise intensity $D$ and input $\omega$ of a single active rotator
for the two potentials $V_\mathrm{cos}$ (a) and $V_\mathrm{opt}$ (b). 
Dashed black lines indicate the transition from excitable to oscillatory regime. Black color on lower panels marks a region with CVs above $1.5$. 
Curves and circles on panels a and b show parametric plots of $(\omega,D)= (\omega_{\text{mod}}(1,N) , D_{\text{mod}}(1,N))$, for the effective noise intensity and the driving parameter of the networked elements obtained from  Eqs. (\ref{eq:modulateddrive}) and (\ref{eq:modulatednoise}). They correspond to the curves depicted in Figs. \ref{stoch-vs-N.fig}b and \ref{stoch-vs-N.fig}c, respectively. 
Arrows indicate the direction of increasing network size $N$. Gray lines and circles refer to curves for excitable and black lines and circles to those for oscillatory peripherals depicted in Fig.~\ref{stoch-vs-N.fig}b,c.
}
  \label{fig:Traces}
\end{figure}

The mean firing rate and the CV of the central node can then be calculated from  the equations for the mean ISI and the ISI variance of a single active rotator, Eqs. (\ref{eq:SelfConstmeanISI}) and (\ref{eq:VarianceStart}), respectively, with modified driving parameter, Eq.~(\ref{eq:modulateddrive}), and noise intensity  Eq.~(\ref{eq:modulatednoise}). 
Consequently, the ISI statistics of the central hub of the star network is that of a single active rotator, Eq.~(\ref{singlerot.eq}), with $\omega = \omega_{\text{mod}}(\timeAv{\rho},N)$ and $D= D_{\text{mod}}(\timeAv{\rho},N)$. 
The statistics for the latter is illustrated in the respective plots in Fig.~\ref{fig:single_rotator} and summarized in Fig.~\ref{fig:Traces} as contour plots.
These theoretical results are compared to simulations in Fig.~\ref{stoch-vs-cpl.fig}b, showing very good correspondence for strong coupling.  
In particular, the theory correctly accounts for the decrease of the firing rate and the increase of the CV in the range of strong coupling, which is also observed in the star network of HH-type nodes. 

Since lower values of the coupling strength enter only through a slightly smaller average degree of synchronization in Eq.~(\ref{eq:ThetaStrongCoupling}), the shaping of the dependence of the firing rate and the CV on $N$ is caused by slightly asynchronous 
behavior of the peripheral nodes. For a fixed number of peripheral nodes this slight asynchrony causes larger tilt $\omega_{\text{mod}}(\timeAv{\rho},N)$ of the corresponding potential and stronger noise $D_{\text{mod}}(\timeAv{\rho},N)$. As can be seen in the contour plots in Fig. \ref{fig:Traces}, this
always causes an increase of the firing rate. However, the CV is reduced only 
if $\omega_{\text{mod}}(\timeAv{\rho},N)<1$ and 
$D_{\text{mod}}(\timeAv{\rho},N)$ is sufficiently small, i.e. when the single active rotator given by Eq. (\ref{eq:ThetaStrongCoupling})  is excitable and below the coherence resonance minimum of the CV.

In the limit of perfectly synchronous spiking of all peripherals, i.e. $\timeAv{\rho} = 1$, the strong coupling approximation, \eq{eq:ArbitraryG}, holds for arbitrary function $G(x)$ and the effective  
parameters $\omega_{\text{mod}}$ and $D_{\text{mod}}$ depend only on the network size $N$ and the nodes' parameters. The related strong coupling approximations for the ISI statistics of the central node for two choices of the potential with $G(x)=-V_{\text{cos}}'(x)$ and $G(x)=- V_{\text{opt}}'(x)$, corresponding simulation results, and simulation results for the HH-type nodes are compared
in Fig. \ref{stoch-vs-N.fig}. As can be seen from the figure, all three models possess qualitatively similar dependencies of the firing rate and the CV vs $N$. 
In the strong coupling regime theory fits perfectly the results from simulations. 

The dependencies $\omega_{\text{mod}}(1,N)$ and $D_{\text{mod}}(1,N)$ can be mapped onto contour maps of the firing rate and the CV of the single rotator as parametric curves, shown in  Fig.~\ref{fig:Traces}.
Indeed, the influence of the number of peripherals $N$ on the ISI statistics can be easily understood by considering the corresponding dependence of the effective parameters  $\omega_{\text{mod}}$ and $D_{\text{mod}}$ on $N$. 
For $N=1$, we find $\omega_{\text{mod}}(1,1)=(\omega_{\theta}+\langle \omega_{\phi}\rangle_N)/2$ and $D_{\text{mod}}(1,1) = (D_{\theta}+\langle D_{\phi}\rangle_N)/4$. In contrast, in the limit of large $N$, we obtain $\omega_{\text{mod}}(1,\infty)=\langle \omega_{\phi}\rangle_N$ and $D_{\text{mod}}(1,\infty) = 0$. Thus, the behavior for large $N$ is determined by the peripheral nodes. Furthermore, an optimal network size, with respect to a minimal CV, can be observed whenever the trace of $(\omega_{\text{mod}}(1,N), D_{\text{mod}}(1,N))$ 
crosses the region of low CVs in Fig. \ref{fig:Traces}. This corresponds to the case when the peripherals are excitable and strong noise is present in the system, i.e. when $D_{\theta}+\langle D_{\phi}\rangle_N$ 
is large.

\section{Conclusion}

We showed that coherent oscillations can emerge in a star network in which peripheral excitable elements receive random and independent inputs and are coupled  via a central excitable hub. The coherence of synchronous stochastic spiking can be maximized by changing the coupling strength between peripheral and central nodes and by the network size, which 
indicates that the observed phenomenon is a specific type of system size  \cite{Pikovsky2002, toral2003system,tessone2007theory} and array-enhanced coherence resonance \cite{NeiSch99, ZhoKur01}. 

We developed an analytical approximation for a generic model of a star network of active rotators. The theory shows that in the strong coupling regime the peripheral nodes can be replaced by an effective rotator describing their mean field phase. 
Using this approach we were able to predict the ISI statistics 
of synchronized oscillations and the optimal network size with respect to a maximal firing rate and a minimal ISI variability. 
This is a specific type of system size resonance \cite{Pikovsky2002, toral2003system,tessone2007theory},  observed in all-to-all coupled noisy elements, whereby 
addition of elements into the network reduces the effective noise in the system. In case of the star network, variation of the network size modifies the excitability of the system, in addition to the noise reduction mentioned above. A reduction of 
the coupling strength leads to an increase of the firing rate and, in certain parameter ranges, to a reduction of ISI variability. The latter leads to array-enhanced coherence resonance, i.e. maximal coherence at a finite coupling strength \cite{NeiSch99, ZhoKur01}. 
According to our analytical results, this is caused by a slight desynchronization of the oscillations of the peripheral nodes. A similar effect has been observed in large networks of globally coupled active rotators \cite{kanamaru2003array, tessone2007theory}.

The emergence of coherent oscillations was investigated using a HH-type node dynamics in which the transition to oscillatory behavior occurs via subcritical Andronov-Hopf bifurcation and in active rotator models in which the transition occurs via a saddle-node bifurcation. This indicates that the observed phenomenon is independent of the respective type of bifurcation and is a generic phenomenon in star networks of excitable elements, similar to emergence of pacemakers in excitable media \cite{jung2000heartbeat}.
Our results suggest a  biologically plausible mechanism for the emergence of noisy periodic discharges observed in branched myelinaed axons of some sensory neurons, such as muscle spindles  \cite{banks1997pacemaker}. In particular, in sensory neurons of cat muscle spindles the length of primary myelinated segments ranges
from 58 to 192 $\mu$m and the distance from the first branching point to 1--9 peripheral nodes ranges over 250 -- 560 $\mu$m \cite{banks1982form}. With such relatively short myelinated segments, electrotonically coupled  nodes are likely to operate in the strong coupling regime studied here.

\section{Acknowledgements}
The authors thank D.F.~Russell and J.T.C.~Schwabedal for valuable discussions. JK and LSG have been funded by IRTG 1740/TRP 2011/50151-0, DFG / FAPESP and by the BMBF (FKZ: 01GQ1001A). AN gratefully acknowledges the support of NVIDIA Corp. with the donation of the Tesla K40 GPU used for this research. 

\appendix
\section{Linear approximation for the phase difference, $\delta^{\theta,\Psi}$}
\label{ap:calculationDelta}

The dynamics of $\delta^{\theta,\Psi}(t)$ is found by subtracting the
individual phase dynamics in \eq{eq:TwoOscillatorRepresentation}. It gives
\begin{align}
  \label{eq:AlphaBeta}
  \dot{\delta}^{\theta,\Psi} = \omega_1-\sin\left( \frac{\beta^{\theta,\Psi}+\delta^{\theta,\Psi}}{2} \right)+\sin\left( \frac{\beta^{\theta,\Psi}-\delta^{\theta,\Psi}}{2} \right) - \kappa_1 \delta^{\theta,\Psi} + \sqrt{2 D_1}\xi_1(t),
\end{align}
with $\beta^{\theta,\Psi}=\theta+\Psi$, the difference of the input parameters 
$\omega_1=\omega_{\theta} - \frac{\langle \omega_{\phi} \rangle_N}{\timeAv{\rho}}$, and the effective coupling
strength $\kappa_1=\kappa \left(N \timeAv{\rho} + 1 \right)$. The
difference of the two noises yields a new Gaussian white noise
$\xi_1(t)$,  with intensity
$D_1=D_{\theta}+\langle D_{\phi} \rangle_N/N \timeAv{\rho}^2$. 
Linearization with respect to $\delta^{\theta,\Psi}$ yields
\begin{align}
\label{eq:AlphaBetaFirstOrder}
\dot{\delta}^{\theta,\Psi} = \omega_1- \left( \cos\left( \frac{\beta^{\theta,\Psi}}{2}
  \right)+ \kappa_1 \right) \delta^{\theta,\Psi} +\sqrt{2 D_1} \xi_1(t).
\end{align}
Furthermore, we can neglect the cosine term inside the brackets since
it is bounded by one and therefore small compared to $\kappa_1$. This yields
\begin{align}
  \label{eq:AlphaOrder}
  \dot{\delta}^{\theta,\Psi} = \omega_1 - \kappa_1 \delta^{\theta,\Psi} +\sqrt{2 D_1} \xi_1(t).
\end{align} 
Thus, $\delta^{\theta,\Psi}(t)$ is an Ornstein-Uhlenbeck process, which  is Gaussian  and can be characterized by a time-dependent mean
and variance. In the strong coupling regime, $\kappa_1$ is large and the mean and variance approach their stationary limits fast.
Then, it is sufficient to approximate $\delta^{\theta,\Psi}(t)$ as the constant mean value $\omega_1/\kappa_1$ plus a white Gaussian noise given by Eq.(\ref{eq:OU}).

\section{Self-consistent calculation of $\timeAv{\rho}$}
\label{ap:calculationRho}

In order to find an approximation for $\timeAv{\rho}$ in Eq. (\ref{eq:ThetaStrongCoupling}), 
we evaluate the time average of the degree of synchronization $\rho$, \eq{eq:timeAvRhoDefinition}. 
Using the definition of the Kuramoto order parameter Eq.
(\ref{eq:KuramotoOrder}) and dividing both sides by $\exp(I \Psi)$, we
find
\begin{eqnarray}
\label{eq:timeAvRho}
 \timeAv{\rho}=\frac{1}{N}\timeAv{\sum \limits_{i=1}^{N} \ e^{I (\phi_i-\Psi)}}= \frac{1}{N}\timeAv{\sum \limits_{i=1}^{N} \ \cos(\phi_i-\Psi)}.
\end{eqnarray}
This follows from the fact that $\rho$ is real and the mean field
phase $\Psi$ is defined in a way that the imaginary part of the sum
vanishes. Next, we introduce the small parameter $\delta^{\theta,\Psi}=\theta-\Psi$. Using a corresponding Taylor expansion and the Kuramoto order parameter Eq.
(\ref{eq:KuramotoOrder}) for the first order term yields  
\begin{eqnarray}
\label{eq:timeAvRhoApprox}
 \timeAv{\rho}\approx \frac{1}{N}\timeAv{\sum \limits_{i=1}^{N} \ \cos(\phi_i-\theta)} + \order{(\delta^{\theta,\Psi})^2}.
\end{eqnarray}
By introducing the conditioned stationary
probability distribution $p(\theta| \timeAv{\rho})$ of $\theta$ for given $\timeAv{\rho}$.
and that of the phases $\phi_i$ of the $i$ peripheral node for a given
particular value of $\theta$, $p_i(\phi_i| \theta)$, we
can apply a similar approach as Tessone et. al. \cite{tessone2007theory} and rewrite Eq.  (\ref{eq:timeAvRhoApprox}) in case of strong coupling into
the self-consistent equation:
\begin{align}
\label{eq:RhoSelfConsistent}
\timeAv{\rho}=\frac{1}{N}\int \limits_{0}^{2 \pi} \ d\theta \ \,
p(\theta| \timeAv{\rho}) \,\sum \limits_{i=1}^N \int
\limits_{0}^{2 \pi} \ d\phi_i \ \cos(\phi_i-\theta) p_i(\phi_i|\theta).
\end{align}

Next, we apply standard methods for stochastic systems \cite{Gar85}, and consider the \textit{Fokker-Planck
  equation} for $p_i(\phi_i| \theta)$ following from the
$\phi_i$-dynamics, Eq. (\ref{eq:NetworkDynamics}) for a constant
$\theta$ first. We are interested in the stationary probability
distribution and have to take into account the instantaneously phase
reset $\phi_i=2 \pi \rightarrow \phi_i=0$. Such a problem corresponds to
the exit problem of a Brownian particle, which starts at $\phi_i=0$ and leaves
the region $(-\infty,2 \pi)$ through the point $\phi_i=2 \pi$ for the
first time.  The \textit{Fokker-Planck equation} for the
stationary probability distribution of this problem reads
\begin{eqnarray}
\label{eq:FPEPhi}
 0 = \partial_t p_i(\phi_i| \theta) =
 \partial_{\phi_i} \left(\partial_{\phi_i} F_i(\phi_i, \theta) + D_{\phi_i} \partial_{\phi_i} \right) p_i(\phi_i| \theta) + r^{\phi_i}(\theta) \delta(\phi_i).\notag \\
\end{eqnarray}
$\delta(\phi_i)$ is the Dirac delta function and $r^{\phi_i}(\theta)$ is the amount of probability that is reset per unit
time and, thus, corresponds to the stationary firing rate of the
system for a given $\theta$.  The drift term is the deterministic flow with $F_i(\phi_i,
\theta)= U(\phi_i) - \kappa_{\phi} \cos(\theta - \phi_i)$ and $U$
defined in \eq{eq:tilded}. Note that driving $\omega_{\phi_i}$ and noise intensity $D_{\phi_i}$ of the individual peripherals differ. Equation (\ref{eq:FPEPhi}) is
complemented by an absorbing boundary $p_i(2 \pi| \theta)=0$ at the
upper end of the region and the probability vanishes as $\phi_i
\rightarrow - \infty$. 
By integrating Eq. (\ref{eq:FPEPhi}) one immediately gets
\begin{align}
 p_i(\phi_i| \theta) \partial_{\phi_i} F_i(\phi_i, \theta) + D_{\phi_i} \partial_{\phi_i}  p_i(\phi_i| \theta) = -r^{\phi_i}(\theta) \ H(\phi_i).
\end{align}
$H(x)$ is the Heaviside step function, which results from the
integration of the delta function in Eq. (\ref{eq:FPEPhi}) and
accounts for the nonzero probability flux between reset and threshold. 
Using variations of constants, this can be integrated and, considering the boundary conditions, we obtain 
\begin{equation}
\label{eq:DistPhi}
p_i(\phi_i| \theta)  = \frac{r^{\phi_i}(\theta)}{D_{\phi_i}} \exp{\left(-\frac{F_i(\phi_i, \theta)}{D_{\phi_i}} \right)} \int \limits_{\phi_i}^{2 \pi} \ d\phi' \ \exp{\left( \frac{F_i(\phi', \theta)}{D_{\phi_i}} \right) }H(\phi').
\end{equation}
Afterwards, the firing rate $r^{\phi_i}(\theta)$ can be obtained
from the normalization condition
resulting in an expression for the firing rate:
\begin{equation}
  \label{eq:rate1}
  r^{\phi_i}(\theta) = D_{\phi_i} \left( \int \limits_{-\infty}^{2 \pi} \ d\phi \ \exp{\left( -\frac{F_i(\phi, \theta)}{D_{\phi_i}} \right) } \int \limits_{\phi}^{2 \pi} \ d\phi' \exp{\left( \frac{F_i(\phi', \theta)}{D_{\phi_i}} \right)}H(\phi') \right)^{-1}.\notag
\end{equation}

One can apply a similar approach for $p(\theta| \timeAv{\rho})$
for a given, fixed value of $\timeAv{\rho}$. To this
end, we approximate the $\theta$-dynamics by the single-node
description, \eq{eq:ThetaStrongCoupling}.  The corresponding
\textit{Fokker-Planck equation} for the stationary distribution
becomes
\begin{eqnarray}
\label{eq:Stationarypthetarho}
  0 &=& \partial_{\theta} \left(\partial_{\theta} U(\theta, \timeAv{\rho}) + D_{\text{mod}} (\timeAv{\rho},N)\partial_{\theta} \right) p(\theta| \timeAv{\rho}) + r^{\theta} (\timeAv{\rho}) \delta(\theta).
\end{eqnarray}
$r^{\theta} (\timeAv{\rho})$ is the stationary firing rate of
the $\theta$-rotator for given $\timeAv{\rho}$.  The potential $U(\theta, \timeAv{\rho})$ depends on the degree of synchronization 
via the tilt $\omega_{\text{mod}}(\timeAv{\rho},N)$ given in
\eq{eq:modulateddrive}.  Additionally, the modified noise
intensity $D_{\text{mod}}(\timeAv{\rho},N)$ is
given in \eq{eq:modulatednoise}. Solving \eq{eq:Stationarypthetarho}, we obtain 
\begin{eqnarray}
  p(\theta| \timeAv{\rho})=\frac{r^{\theta} (\timeAv{\rho})}{D_{\text{mod}}(\timeAv{\rho},N)} \exp{\left(-\frac{V(\theta, \timeAv{\rho})}{D_{\text{mod}}(\timeAv{\rho},N)} \right)} \int \limits_{\theta}^{2 \pi} \ d\theta' \exp{\left( \frac{V(\theta',\timeAv{\rho})}{D_{\text{mod}}(\timeAv{\rho},N)} \right)}H(\theta').
\end{eqnarray}
Again, the firing rate $r^{\theta} (\timeAv{\rho})$ can be obtained from the normalization condition.

Using the results for $p(\theta| \timeAv{\rho})$ and $p(\phi_i| \theta)$, we can calculate the right-hand side
of the self-consistent Eq. (\ref{eq:RhoSelfConsistent}) numerically.
For the evaluation, we fix $\timeAv{\rho}$ and calculate the
corresponding right-hand side.  Continuing this for $\timeAv{\rho}
\in [0,1]$, we search for intersections of the left-hand and
right-hand side.  For all considered parameter sets, we
observed only one intersection.  Implementing this into the single-node 
 representation, Eq. (\ref{eq:ThetaStrongCoupling})
, yields a
full strong coupling approximation of the dynamics, which is used in Fig. \ref{stoch-vs-cpl.fig}b.

%

\end{document}